\documentclass[prd, aps, superscriptaddress, twocolumn, preprintnumbers, floatfix, nofootinbib]{revtex4}
\usepackage[dvips]{graphicx}
\usepackage{epsf}
\usepackage{amsmath}
\usepackage{amssymb}

\usepackage{graphicx}
\usepackage{dcolumn}
\usepackage{bm}

\def\be{\begin{equation}}
\def\ee{\end{equation}}
\def\bea{\begin{eqnarray}}
\def\eea{\end{eqnarray}}

\usepackage{color}

\begin{document}


\title{A Note on Trans-Planckian Tail Effects}
\author{L.L. Graef$^{1,2}$ and R. Brandenberger$^{}$}
\email{leilagraef@usp.br, rhb@physics.mcgill.ca}

\affiliation{Department of Physics, McGill University, 
Montr\'eal, QC, H3A 2T8, Canada \\
$^{2}$ Instituto de F\'isica, Universidade de S\~ao Paulo, 
Rua do Mat\~ao travessa R, 05508-090, S\~ao Paulo, SP, Brazil}


\begin{abstract}

We study the proposal by Mersini et al \cite{Mersini} that 
the observed dark energy might be explained by the
back-reaction of the set of {\it tail modes} in a theory with
a dispersion relation in which the mode frequency decays
exponentially in the trans-Planckian regime.  
 The matter tail modes are
frozen out, however they induce metric fluctuations.
The energy-momentum tensor with which the tail modes
effect the background geometry obtains contributions
from both metric and matter fluctuations. 
We calculate the equation of state induced by the tail modes 
taking into account the gravitational contribution.
We find that,
in contrast to the case of frozen super-Hubble cosmological
fluctuations, in this case the matter perturbations
dominate, and they yield an equation of state which to
leading order takes the form of a positive cosmological
constant.
 
\end{abstract}

\maketitle

\section{Introduction}

In an interesting paper, Mersini et al \cite{Mersini}
have put forwards the suggestion that the energy density
stored in trans-Planckian modes provides a candidate
for the observed dark energy of the universe. The
authors of \cite{Mersini} assumed that the dispersion
relation for the fluctuation modes of some matter field
$\varphi$ is dramatically
modified for wave numbers larger than some critical
value $k_c$. If we think in terms of waves propagating in an inhomogeneous medium, it is
reasonable to assume that the dispersion relation for the mode propagation will
get modified when the mode start probing the underlying structure of the background,
in our case, the trans-Planckian regime \cite{gil}. Instead of the usual linear dispersion
relation for the frequency $\omega_k$, namely $\omega_k \sim k$, 
a decaying function of Epstein class \cite{Epstein} of the form
\be
\omega_k^2 \, = \, k^2 \bigl( \frac{\epsilon_1}{1 + e^x} + 
\frac{\epsilon_3 e^x}{(1 + e^x)^2} \bigr) \, ,
\ee
with 
\be
x \, \equiv \bigl( \frac{k}{k_c} \bigr) 
\ee
was considered. Here, $\epsilon_1$ and $\epsilon_3$ are
constants. The dispersion function above encapsulates the T-duality behaviour. 
To obtain the linear dispersion relation for
$k \ll k_c$, the values $\epsilon_1$ and $\epsilon_3$ must
satisfy the constraint
\be
\frac{\epsilon_1}{2} + \frac{\epsilon_3}{4} \, = \, 1.
\ee
The value of $k_c$ is given by
the high energy scale of the new physics (e.g. string
physics) which leads to the change in the dispersion
relation.

The family of dispersion models chosen attenuates to zero in the 
trans-Planckian regime, thereby producing ultralow frequencies at 
very short distances.  The total energy contribution of the modes
produced is then finite, eliminating the need for renormalization. 

The modes with very high momenta but ultralow frequencies $w(k)$ 
are frozen for as long as the Hubble expansion rate of the universe 
dominates over their frequencies. These modes correspond to the 
ones with  $k > k_H$ where $k_H$ is determined via
\be
\omega^2(k_H) \, = H^2,
\ee
$H$ being the Hubble expansion rate. 
In \cite{Mersini},\cite{gilmersini} these are called the {\it tail modes}. 
It then follows that the tail modes have not decayed and redshifted 
away but are still frozen today.

The energy density in the tail modes can be computed
in the standard way \cite{Mersini, BD}
\begin{equation}
<\rho_{tail}> \, = \, \frac{1}{2\pi^{2}}\int_{k_{H}}^{\infty} k dk \int \omega_k d\omega |\beta_{k}|^{2},
\end{equation}
where $\beta_k$ are the Bogoliubov coefficients which
give the excitation level of the mode. These
coefficients were calculated in \cite{Mersini} assuming
the modes begin in the state which minimizes the
Hamiltonian (the same prescription as the one used
in \cite{MB}). Not unexpectedly, it was found that these
coefficients decay exponentially as $k \rightarrow \infty$.
Rather surprisingly, it was found then that
$<\rho_{tail}>$ is comparable in amplitude to the
observed dark energy density, without any fine-tunning and 
with the Planck mass $m_{pl}$ being the fundamental 
scale $k_c$ of the theory.

In this note we take a closer look at this suggestion.
It is not sufficient to simply calculate the energy
density stored in the tail modes, but we also have to
verify that their equation of state is really that of dark 
energy. In addition, we must consider the effects of
the gravitational fluctuations which are induced by
the matter fluctuations. As studied in \cite{ABM},
the effect of the matter and metric fluctuations
on the background cosmology can be described in terms
of an {\it effective energy momentum tensor $\tau_{\mu \nu}$}.
Here, we study the equation of state of $\tau_{\mu \nu}$ for the tail modes
and we find that, given certain assumptions, it indeed
has the right form to be a candidate for dark energy.

In the following, we first review the effective energy
momentum tensor $\tau_{\mu \nu}$ of cosmological fluctuations. In 
Section 3 we then study the equation of state of the tail
modes contribution to $\tau_{\mu \nu}$. We end with a discussion of
our results.
Throughout this paper we will use natural units in which the speed of light
and Planck's constant are set to $1$. Greek letters are used
for space-time indices and latin indices run over spatial dimensions
only.   
\\

\section{Effective Energy Momentum Tensor for Cosmological Fluctuations}

As is well known, cosmological fluctuations are observed to be small
today on large scales (scales which we see in the cosmic
microwave background), and hence a linear analysis of
the fluctuations on top of the evolving background
cosmology is usually a good starting point. Linear fluctuations can be analyzed in Fourier space where each
Fourier mode evolves independently. The Einstein field equations, however, are highly nonlinear and even at the classical level fluctuations at second order influence the background. So, in this paper we
will be concerned with the leading effects which the
fluctuations have on the background beyond the linear
treatment. At quadratic order
there is a coupling between the Fourier modes. In particular,
a Fourier mode with wave vector ${\bf k}$ can combine with
a mode with wave vector ${\bf -k}$ to yield a correction to the
background, which is the ${\bf k = 0}$ mode. This effect is
called {\it back-reaction}.

On the other hand, matter fluctuations induce, via the Einstein
constraint equations, metric fluctuations. These metric fluctuations
will also back-react on the homogeneous background metric. The
evolution of the full metric is governed by the Einstein equations
\be \label{Einstein}
G_{\mu \nu} \, = \, 8 \pi G T_{\mu \nu} \, ,
\ee
where $G_{\mu \nu}$ is the Einstein tensor of the metric $g_{\mu \nu}$,
$T_{\mu \nu}$ is the energy-momentum tensor of matter, and $G$ is
Newton's gravitational constant. 

At linear order in the amplitude
of cosmological fluctuations, the metric and matter can be written
as
\be \label{metric}
ds^2 \, = \, (1 + 2 \phi({\bf{x}},t)) dt^2 - 
a(t)^2 \bigl[ (1 - 2 \psi({\bf{x}},t)) \gamma_{ij}dx^i dx^j \bigr] \, ,
\ee
and
\be \label{matter}
\varphi({\bf{x}},t) \, = \, \varphi_o(t) + \delta \varphi({\bf{x}},t) \, .
\ee 
We have chosen a particular 
coordinate system (longitudinal gauge) in order to write the metric
in the form (\ref{metric}) \footnote{See \cite{MFB} for an in-depth
review of the theory of cosmological fluctuations and \cite{RHBfluctrev}.
for an introductory overview.} In the longitudinal gauge  the metric 
fluctuations $\phi$ and $\psi$  coincide with Bardeen's gauge 
invariant variables $\Phi$ and $\Psi$. In the absence of 
anisotropic stress $\phi$ and $\psi$ are equal. In (\ref{metric}),
$\gamma_{ij}$ is the background metric of the constant time hypersurfaces.
For vanishing spatial curvature (the case we will discuss) we simply
have $\gamma_{ij} = \delta_{ij}$. The background metric is given
by the scale factor $a(t)$, the background matter by $\varphi_0(t)$. 
We will only consider scalar metric fluctuations. 
Vector perturbations decay in an expanding universe, and since they
are not seeded at linear order by matter fluctuations we can neglect them.
In the next section we will also comment on the possible role
of gravitational waves.

To find out how the linear fluctuations effect the background (an
effect which is quadratic in the amplitude of the inhomogeneities)
we insert (\ref{metric}) and (\ref{matter}) into the Einstein
equations (\ref{Einstein}) and expand to second order. The zero'th
order terms obey the background equations, the linear terms
are assumed to satisfy the linear perturbation equations. The
terms on the left hand side of (\ref{Einstein}) which
are quadratic can be moved to the right hand side of the equation,
where they combine with the quadratic terms in $T_{\mu \nu}$ to
form an {\it effective energy-momentum tensor}
\be \label{effT}
\tau_{\mu \nu}({\bf x},t) \, 
\equiv \, T^{(2)}_{\mu \nu} - \frac{1}{8 \pi G} G^{(2)}_{\mu \nu}
\, ,
\ee
where the superscript (2) indicates the order of the terms. 

In the presence of fluctuations, the background metric is modified.
The corrected background is
\be
g^{(bg)}_{\mu \nu}(t) \, \equiv \, g^{(0)}_{\mu \nu}(t) + \delta g^{(bg,2)}(t) 
\, ,
\ee
where the first term is the background metric and the second one
indicates the quadratic corrections to the background. Following
the method proposed in \cite{ABM} and reviewed in \cite{RHBBRrev},
we can extract the corrections to the background by taking the
spatial average of (\ref{Einstein}) expanded to second order.
The equation then takes the form
\be \label{EinsteinBR}
G_{\mu \nu}(g_{\alpha \beta}^{(bg)}) \, = \, 8 \pi G \tau_{\mu \nu}(t) \, ,
\ee
where $\tau_{\mu \nu}(t)$ is the spatial average of (\ref{effT}).

The form of the effective energy-momentum tensor $\tau_{\mu \nu}$ was
derived in \cite{ABM} with the result
\begin{align}\label{tau00}
\tau_{00} \, = \, & \frac{1}{8\pi G}[+12H<\phi \dot{\phi}> - 3<(\dot{\phi})^{2}> + 9a^{-2}<(\nabla \phi)^{2}>]
\nonumber \\ 
& + \frac{1}{2}<(\delta \dot{\varphi})^{2}> + \frac{1}{2} a^{-2}<(\nabla \delta \varphi)^{2}> \\
& + \frac{1}{2} V"(\varphi_{0})<\delta \varphi^{2}> + 2V'(\varphi_{0})<\phi \delta \varphi> \nonumber
\end{align}
and
\begin{align}\label{tauij}
\tau_{ij} \, = \, & a^{2} \delta_{ij} \{\frac{1}{8\pi G}[(24H^{2}+16\dot{H})<\phi^{2}> + 24 H<\dot{\phi} \phi> 
\nonumber \\
& + <(\dot{\phi})^{2}> + 4<\phi \ddot{\phi}>  -\frac{4}{3} a^{-2}<(\nabla \phi)^{2}>] \nonumber \\ 
& + 4\dot{\varphi_{0}^{2}}<\phi^{2}> + \frac{1}{2}<(\delta \dot{\varphi})^{2}> \\ 
& - \frac{1}{6} a^{-2}<(\nabla \delta \varphi)^{2}> - 4 \dot{\varphi_{0}}<\delta \dot{\varphi} \phi>  \nonumber \\ 
& - \frac{1}{2} V"(\varphi_{0})<\delta \varphi^{2}> + 2V'(\varphi_{0})<\phi \delta \varphi>\}. \nonumber
\end{align}
Inserting these expressions into (\ref{EinsteinBR}) allows the determination of the
effect of cosmological perturbations on the background metric.

In \cite{ABM},\cite{martineau},\cite{renato} the backreaction of long wavelength modes was studied. In \cite{ABM} it was found that the contribution of super-Hubble modes to $\tau_{\mu \nu}$
acts like a negative cosmological constant, and hence can potentially lead to a
dynamical relaxation of the cosmological constant, as discussed in \cite{RHBBRrev}.
The effect is easy to understand heuristically: for super-Hubble modes the contribution
of spatial gradients is negligible. Since the dominant mode of $\phi$ is constant in
time on super-Hubble scales, if the equation of state of matter is constant then time
derivative terms are also negligible. Thus, the equation of state of $\tau_{\mu \nu}$
has to be that of a cosmological constant. Matter fluctuations carry positive energy,
but lead to metric potential wells which have negative gravitational energy. On
super-Hubble scales the negativity of the gravitational energy overcomes the
positivity of the matter energy, hence explaining the sign of the effect.

An important issue first raised in \cite{Unruh} is whether the resulting correction to the 
background metric is physical, or whether it is equivalent to a second order time
reparametrization. In fact, in the case of long wavelength (super-Hubble) adiabatic fluctuations
it can be shown \cite{Ghazal1, Abramo} that the effect is indeed not physically measurable. 
The local expansion rate of space computed at a fixed value of the only
clock field in the problem, the matter field $\varphi$, is independent of whether there
are fluctuations $\phi$ or not. However, in our universe there are several matter fields.
In particular, we measure time in terms of the temperature of the CMB, i.e. in terms
of a clock field which has a negligible effect on the expansion of space.
 Since we are
interested in effects in the current universe, we can assume that time is measured
from the CMB and thus back-reaction effects computed via (\ref{EinsteinBR})
are physical \cite{Ghazal2}. This is similar to the case discussed in \cite{Marozzi} where it was shown that the backreaction of long wavelengths fluctuations is physical and leads in fact to a decrease in the cosmological constant. 
\\

\section{Tail Mode Contribution}

We now want to study the contribution of tail modes to $\tau_{\mu \nu}$
in the model of \cite{Mersini}, where the dispersion relation is dramatically
modified in the ultraviolet. The modification of the dispersion relation can be
taken into account by replacing the $\nabla$ operator in Fourier space
by $\omega_k \nabla$ in the relations (\ref{tau00}) and (\ref{tauij}).

Whereas for the standard dispersion relation the contribution of short
wavelength modes to $\tau_{\mu \nu}$ is divergent, it is convergent
in our case. 
Since the dynamics of the tail modes is frozen (no oscillations
on the time scale $k^{-1}$) we might expect that, in analogy to the
case of the frozen super-Hubble modes, the contribution of the
tail modes might look like a cosmological constant, and might hence
contribute to dark energy. Here we study this question.

Matter and metric fluctuations $\delta \varphi$ and $\phi$ are not independent. They are
related through the $0i$ component of the Einstein equations,
\begin{equation} \label{EinsteinCE}
\dot{\phi} + H \phi \, = \, 4\pi G \dot{\varphi_{0}} \delta \varphi.
\end{equation}
Note that if we were to use this equation to determine $\delta \varphi$
from $\phi$, there would be a singularity when $\dot{\varphi_{0}}=0$. This
corresponds to the breakdown of longitudinal gauge. However, we will be
using this equation to determine the metric fluctuations from the matter
inhomogeneities, and thus no problem in applying (\ref{EinsteinCE})
arises.  

The above equation has a homogeneous solution, which is
constant modulo Hubble damping, and an inhomogeneous
solution determined by the matter fluctuations,
which to leading order in $H$ obeys the equation
\be \label{EE}
\dot{\phi} \, \approx \, 4 \pi G \dot{\varphi_0} \delta \varphi \, .
\ee

We study back-reaction in the matter-dominated phase of Standard
Big Bang cosmology. We will use the scalar field $\varphi$ to model
pressureless matter.  A massive scalar
field with potential
$V(\varphi) = m^2 \varphi^2/2$
has an equation of state whose time average can yield
pressureless matter. The mass $m$ sets the time scale on which the
equation of state oscillates about $p = 0$, where $p$ denotes
the pressure. This time scale should be microscopic in order not
to lead to cosmologically relevant effects. Hence, we require
$m \gg H$. 

The oscillations of the matter fluctuations corresponding to the tail modes, which we are considering, are prevented due to the modified dispersion relation chosen. Hence, from equation (\ref{EinsteinCE}) we see that the oscillations of $\phi$ are actually driven by the oscillating background. Therefore, the fast oscillating quantities are  $\varphi_0(t)$, ${\dot \varphi_0}(t)$, and $\phi$. If we are interested in the time-averaged equation of state
of $\tau_{\mu \nu}$, we can drop all terms which are linear
in rapidly oscillating functions. 

Hence, if $\varphi_0(t)$ is oscillating as
\be
\varphi_0(t) \, = \, {\cal A} {\rm sin}(mt), 
\ee
then $\phi$ will oscillate as
\be
\phi \, = \frac{1}{m} {\tilde{\cal A}} {\rm sin}(mt), \,
\ee
with space-dependent amplitude ${\tilde{\cal A}}$ determined
by the space-dependent matter fluctuation via
\be \label{amplrel}
{\tilde{\cal A}} \, = \, 4 \pi G m {\cal A} \delta \varphi \, .
\ee

In order for $\varphi$ to dominate the energy density
we have (from the Friedmann equation)
\be
m^2 {\cal A}^2 \, \approx \, m_{pl}^2 H^2 \, ,
\ee
where $m_{pl}$ is the reduced Planck mass and ${\cal A}$ is the amplitude of the scalar field. Hence
\be \label{A}
{\cal A} \, \approx \, \frac{H}{m} m_{pl} \, \ll \, m_{pl} \, .
\ee

In the following we will use these results to estimate the magnitude
of the tail mode contribution to $\tau_{\mu \nu}$.

We first compare the magnitude of the gravitational terms
with those of the matter terms. The matter terms are
\be
\tau_{00}^m = \frac{1}{2}\left[<(\delta \dot{\varphi})^{2}> + a^{-2}<(\nabla \delta \varphi)^{2}>  
+ V"(\varphi_{0})<\delta \varphi^{2}>\right]
\ee
and
\be
\tau_{ii}^m=\frac{1}{2}<(\delta \dot{\varphi})^{2}>  - \frac{a^{-2}}{6}<(\nabla \delta \varphi)^{2}> 
- \frac{V"(\varphi_{0})}{2} <\delta \varphi^{2}>,
\ee
where we recall that the $\nabla$ operator is the modified one.
Since the tail modes are frozen and the $\nabla$ operator is the
modified one, the first two terms on the right hand side of each
equation are suppressed, and the third term dominates. Hence,
the induced equation of state of these matter terms is
\be
p \, \simeq \, - \rho \, ,
\ee
where $\rho$ is the energy density. The correction terms
coming from the first two terms on the right hand side of
each equation leads to
\be
w \, \equiv \, \frac{p}{\rho} \, > \, -1 \, .
\ee

Now let us turn to the terms which are quadratic in the gravitational
potential $\phi$. Since the rate of change of $\phi$ has magnitude
$m \phi$, we can neglect the terms proportional to $H$ and $\dot{H}$ (and consequently also $\dot{\varphi}_{0}^{2}$)
in (\ref{tau00}) and (\ref{tauij}). So we obtain

\be \label{rhograv}
\tau_{00}^g \, \approx \, - 3 m_{pl}^2 m^2 < \phi^2 > + 9 m_{pl}^{2} a^{-2}<(\nabla \phi)^{2}>
\ee
and
\be \label{pgrav}
\tau_{ii}^g \, \approx \, - 3 m_{pl}^2 m^2 < \phi^2 > -\frac{4}{3} m_{pl}^{2} a^{-2}<(\nabla \phi)^{2}> \, .
\ee

Now let us estimate the magnitude of the matter term $<(\delta \varphi)^2>$ in order 
to compare the matter and the gravitational contributions.
From eq. (\ref{EE}) we have that $m_{pl}^{2} \phi \approx \varphi_{0} \delta\varphi$, 
since both $\phi$ and $\varphi_{0}$ are fast oscillating quantities. Then we have 
\begin{equation}
\delta\varphi^{2}  \approx  \frac{m_{pl}^{4} \phi^{2}}{{\cal A}^{2}}. 
\end{equation}
Substituting ${\cal A}$ from eq. (\ref{A}) and considering that $\tau_{00}^{m} \approx m^{2}<\delta\varphi^{2}>$, we obtain 
\begin{equation}
\tau_{00}^{m} \approx \frac{m^{4}m_{pl}^{2} <\phi^{2}>}{H^{2}} = m^{2} m_{pl}^{2} <\phi^{2}> \left(\frac{m^{2}}{H^{2}}\right).
\end{equation}
Comparing the above equation  with $\tau_{00}^{g}$, we can see that $\tau^{m}_{00} \approx (m^{2}/H^{2})\tau^{g}_{00}$. 
Since $m>>H$, we conclude that the
matter contribution to the effective energy-momentum tensor dominates over the 
gravitational contribution. 

The time average of the cross terms (the terms involving one factor
of $\phi$ and one factor of $\delta \varphi$ )vanishes and hence
we can neglect these terms. Thus, in summary, we find that
the matter terms in the effective energy-momentum tensor
$\tau_{\mu \nu}$ dominate. They lead to an equation of state
$w = -1$ with positive effective energy density. Therefore, it appears
that, even when taking gravitational effects into account, the tail
modes can provide a candidate for dark energy.

For completeness, we are now going to estimate the effective energy-momentum 
tensor of the gravitational waves  for this class of dispersion relation.

In the case of gravitational waves the metric assumes the form
\begin{equation}
ds^{2} \, = \, dt^{2} - a^{2}(t)(\delta_{ik} + h_{ik})dx^{i} dx^{k},
\end{equation}
where $h_{ik}$ is defined as the transverse traceless part of the metric perturbations.

In \cite{ABM} it was shown that, neglecting the coupling of the gravitational waves 
to matter, the effective energy momentum tensor of gravitational waves is given by
\bea
8\pi G \tau_{0}^{0} \, &=& \,  \frac{\dot{a}}{a} <\dot{h}_{kl} h_{kl}> \\
& & \, + \frac{1}{8} \left(<\dot{h}_{kl}\dot{h}_{kl}> + \frac{1}{a^{2}}<h_{kl,m} h_{kl,m}>\right)
\nonumber 
\eea
and
\begin{equation}
- \frac{8\pi G}{3} \tau^{i}_{i} \, = \, \frac{7}{24 a^{2}} <h_{kl,m} h_{kl,m}> 
- \frac{5}{24} <\dot{h}_{kl} \dot{h}_{kl}>,
\end{equation}
where we assumed that the gravitational wave field is isotropic.

In the tensorial case, the quantity that can be interpreted as the pressure is \cite{ABM}
\begin{equation}
p_{gw} \, = \, -\frac{1}{3} \tau^{i}_{i} - \frac{1}{6H} <\dot{h}^{ij} h_{ij}>(\rho_{0} + p_{0}),
\end{equation}
while the energy density is still given by $\rho_{gw}= \tau^{0}_{0}$. 

Since we are interested in the tail modes, which are frozen, we can neglect all 
terms with either space or time derivatives. Thus,
all terms are negligible, and we conclude that the tail modes
of the spectrum of gravitational waves only have a negligible
effect on the effective energy-momentum tensor of back-reaction.

\section{Discussion}

We have studied the equation of state of the effective
energy-momentum tensor with which tail modes back react on
the background space-time. The context of our study is
a proposal by Mersini et al. \cite{Mersini} that the dispersion
relation of fluctuations might be highly distorted in the trans-Planckian 
regime, such that for very large values of $k$
the effective frequency $\omega_k$ becomes smaller than the
Hubble expansion rate $H$ ({\it tail modes}). This class of dispersion 
relation reflects the T-duality symmetry of string theory and leads 
naturally to a finite vacuum energy. It was found in \cite{Mersini} that the 
energy density stored in the tail modes coincides with the energy density 
of the observed dark energy. So it was suggested that, with the correct 
equation of state, this contribution could account for the observed dark 
energy without any fine-tuning. 

In order to verify if this effect could really represent dark energy, we 
have explicitly calculated the equation of state induced by the tail modes, 
including not only the contribution of the matter fluctuations but also that 
of the metric fluctuations. We
have shown that (in contrast to what happens for super-Hubble
modes in the case of a standard dispersion relation)
the effective energy-momentum tensor is dominated by the contributions 
from the matter terms which, to first approximation, leads to an equation of state
of dark energy with positive energy density.
We also analysed the contribution of gravitational 
waves in this regime and found that it is negligible.

In conclusion, our results support the suggestion that the backreaction 
of the trans-Planckian modes might account for the observed dark energy.

\acknowledgements{The authors wish to thank E.G.M. Ferreira and E.L.D. Perico for the useful discussions. RB is supported by an NSERC Discovery Grant, and by 
funds from the Canada Research Chair program. LG is supported by FAPESP 
under grants 2012/09380-8 .}

\end{document}